\input{aipcheck}

\documentclass[
    ,final            
  ]
  {aipproc}

\layoutstyle{6x9}

\begin{document}
\begin{flushright}
CERN-PH-TH/2005-122
\end{flushright}

\title{Heavy flavours: theory summary\footnote{Talk 
given at DIS 2005, XIII Workshop on Deep Inelastic Scattering,
April 27--May 1, 2005, Madison, WI, U.\ S.\ A.}}


\author{G. Corcella}{
  address={CERN, Department of Physics, 
Theory Division, CH-1211 Geneva 23, Switzerland}
}

\begin{abstract}
I summarize the theory talks given 
in the Heavy Flavours Working Group.
In particular, I discuss heavy-flavour
parton distribution functions, threshold resummation for heavy-quark
production, progress in fragmentation functions, quarkonium production,
heavy-meson hadroproduction.   
\end{abstract}

\maketitle



\section{Introduction}

Heavy-flavour physics is currently one of the main fields of investigation,
in both theoretical and experimental particle physics.
In this talk, I summarize the main theoretical issues that were 
presented in the Heavy Flavour session of DIS 2005.
Among the topics discussed, we had updates on heavy-flavour parton 
distributions, large-$x$ resummation for heavy quark production in DIS, 
heavy-quark fragmentation functions to next-to-next-to-leading
order, progress in heavy-meson and quarkonium production.

\section{Heavy-quark parton distributions} 

Relevant work was carried out on the subject of heavy-quark
parton distribution functions. Progress was reported by R.~Thorne, from 
the MRST collaboration, concerning
the formulation of the Variable Flavour
Number Scheme (VFNS) for heavy quarks at NNLO.

In fact, there are problems in defining VFNS for heavy quarks. 
When switching naively from a given order in the coupling
constant $\alpha_S$ with $n_f$ flavours, to the same
order, but with $n_f+1$ flavours, one would get a discontinuity
when $Q^2$ is equal to the heavy quark mass. 
At NNLO, for example, the contribution to a heavy-quark structure function
$F_2^H(x,Q^2)$ with $n_f$ flavours is 
$\sim\alpha_S^3C_2^{\mathrm{FF},3}\otimes f^{n_f}$, where 
$C_2^{\mathrm{FF},3}$ is the
fixed-flavour (FF) coefficient function
and $f$ the parton distribution function.
The corresponding contribution, in the region of
$n_f+1$ flavours, is instead 
$\sim\alpha_S^2C_2^{\mathrm{VF},2}\otimes f^{n_f+1}$, where
$C_2^{\mathrm{VF},2}$ is the variable-flavour (VF)
coefficient function. As a result, $F_2^H(x,Q^2)$ would be
 discontinuous through the heavy-quark mass threshold $Q^2=m_H^2$.

The Thorne--Roberts (TR) prescription handles this discontinuity by freezing
higher-order terms when crossing the value $Q^2=m_H^2$.
At LO, for instance, the TR prescription reads, in terms of gluon ($g$) and
heavy quark ($h$, $\bar h$) densities:
\begin{eqnarray}
{{\alpha_S(Q^2)}\over{4\pi}}C_{2,Hg}^{\mathrm{FF},1}
\left({{Q^2}\over{m_H^2}}\right)\otimes g^{n_f}(Q^2)&\to &
{{\alpha_S(M^2)}\over{4\pi}}C_{2,Hg}^{\mathrm{FF},1}(1)\otimes g^{n_f}(m_H^2)
\nonumber\\
&+&
C_{2,HH}^{\mathrm{VF},0}\left({{Q^2}\over{m_H^2}}\right)\otimes 
(h+\bar h)(Q^2).
\label{tr}
\end{eqnarray}
In Eq.~(\ref{tr}), we note the `frozen' term 
$\sim\alpha_S(M^2)C_{2,Hg}^{\mathrm{FF},1}(1)$.
In order to apply this prescription to NNLO, 
we would need the FF ${\cal O}(\alpha_S^3)$
heavy-quark coefficient functions for $Q^2\leq m_H^2$, 
which have not been computed yet.
However, a reliable approximation of such functions can be obtained, gaining
information from the available calculations which resum threshold logarithms 
and leading $\ln(1/x)$ terms, and  from the known NNLO
coefficient functions and space-like splitting functions.

Using this method, parton densities are still discontinuous at
$Q^2=m_H^2$, but structure functions are now continuous. 
Results were presented for the charm-quark 
structure function $F_2^c(x,Q^2)$: the
NNLO VFNS prediction fits rather well the ZEUS and H1 data, while the NLO
result is always below the data.

F.~Olness, from the CTEQ collaboration, also 
presented recent work on heavy-quark parton
distributions, implemented in the CTEQ6HQ set. 
Mainly, the new set resums logarithms of the heavy-quark mass
$\ln(m_H^2/\mu_F^2)$ in the full kinematic range, by reabsorbing them
in the heavy-quark distribution function. 
This yields a small, but
visible difference with respect to the previous set CTEQ6M.
The predictions obtained using the CTEQ6HQ parton distributions
fit the HERA data on $F_2^c(x,Q^2)$ quite well.

Progress was reported on the determination of the strange-quark density,
which, in the previous CTEQ sets,
was tied to the $\bar u$ and $\bar d$ distributions
via the relation $s=\bar s=\kappa(\bar u+\bar d)/2$. 
This assumption might lead to an underestimate of the 
uncertainty on the $s$-quark density. 
Rather, the $s$ quark can be treated as an additional set, described
by a more
general parametrization, which can be fitted to data.
Preliminary results with an independent $s$-quark density give reasonable
$\chi^2$ when comparing with DIS and Tevatron data.

Moreover, the talk by F.~Olness also discussed soft-resummation for
$b$ production in DIS, which might also be included in future fits
of CTEQ parton distribution functions.

\section{Threshold resummation}\vspace{-0.45cm}
\section{for heavy-quark production in DIS}

Soft-gluon resummation for heavy-quark production in DIS was 
discussed in the talk by A.~Mitov. In fact, the DIS $\overline{\mathrm{MS}}$
coefficient functions present, at ${\cal O}(\alpha_S)$, 
terms $\sim 1/(1-x)_+$ and $\sim [\ln(1-x)/(1-x)]_+$, which become large
for $x\to 1$ and need to be resummed (threshold resummation).
Such contributions correspond to collinear- or soft-gluon radiation.

It was pointed out that the large-$x$ 
behaviour of the coefficient functions crucially depend, via the ratio
$m/Q$, on the mass of the final-state quark. In fact, a light quark emits
both soft- and collinear-divergent radiation, while gluon radiation off
a heavy quark can only be soft-enhanced. As a result, the two regimes
$m/Q\simeq 1$ and $m/Q\ll 1$ must be treated separately when performing
large-$x$ resummation.

The threshold resummation reported by A.~Mitov
was performed in the next-to-leading
logarithmic approximation (NLL), which corresponds to keeping in the
Sudakov exponent terms $\sim\alpha_S^n\ln^{n+1} N$ (LL) and
$ \sim\alpha_S^n\ln^n N$ (NLL), where $N$ is the Mellin moment variable.
In fact, soft resummation is analytically performed in $N$-space, and the
results are then inverted to $x$-space.
Predictions were presented for the charm-quark structure function 
$F_2^c(x,Q^2)$, for charged-current interactions, in
the environment of the HERA and NuTeV
experiments. The results showed that threshold resummation
is relevant, especially at small $Q^2$. Moreover, resumming large-$x$ terms
in the $\overline{\mathrm{MS}}$ coefficient function yields a milder
dependence on both factorization and renormalization scales,
which corresponds to a reduction of the theoretical uncertainty.

\section{NNLO perturbative fragmentation function}

A.~Mitov also reported on 
progress in heavy-quark fragmentation functions.
The energy distribution of a heavy quark presents, at fixed
order, terms $\sim\alpha_S^n\ln^k(Q^2/m^2)$ ($k\leq n$), 
where $Q$ is the hard scale of the process, that are large for
$m\ll Q$, which is often the case.
Such logarithms can be resummed using
the approach of perturbative fragmentation functions, which
expresses the energy spectrum of a heavy quark as the 
convolution of a coefficient function, describing the emission of a 
massless parton, and a perturbative fragmentation function $D(m,\mu_F)$,
associated with the fragmentation of a massless parton 
(a light quark or a gluon) into a massive quark. 
The dependence of $D(m,\mu_F)$ on the factorization scale $\mu_F$
is determined by solving the Dokshitzer--Gribov--Altarelli--Parisi
(DGLAP) evolution equations, once an initial condition at a scale
$\mu_{0F}$ is given. Neglecting powers $(m/Q)^p$,
the initial condition of the perturbative fragmentation function was proved 
to be process independent and calculated several years ago to NLO.
This talk discussed the recent calculation of 
NNLO contributions, i.e. up to ${\cal O}(\alpha_S^2)$.

Denoting by 
$Q$ and $q$ heavy and light quarks respectively, NNLO corrections
to the initial condition of the perturbative fragmentation function
come from the elementary processes: 
$Q\to Qgg$, $Q\to Qq\bar q$, $Q\to QQ\bar Q$, $\bar Q\to Q\bar Q \bar Q$,
$q(\bar q)\to Q\bar Q q(\bar q)$ and $g\to Q\bar Q g$, which have now
been fully computed.

Solving the DGLAP equations, for an evolution from $\mu_{0F}$ to $\mu_F$,
allows us to resum terms $\sim\ln(\mu_F^2/\mu^2_{0F})$, i.e. 
$\sim\ln(Q^2/m^2)$ if we choose $\mu_{0F}\simeq m$ and $\mu_F\simeq Q$
(collinear resummation).
In particular, the leading logarithms 
are $\alpha_S^n\ln^n(Q^2/m^2)$, the NLLs 
$\alpha_S^n\ln^{n-1}(Q^2/m^2)$, the NNLLs $\alpha_S^n\ln^{n-2}(Q^2/m^2)$.
In principle, the calculation of the initial condition of the perturbative
fragmentation function to NNLO would allow one to study the spectrum of heavy
quarks in the NNLO approximation, with NNLL collinear resummation.
However, for this level of accuracy to be achieved, one would  also need
NNLO Altarelli--Parisi time-like splitting functions, which are
currently known to NLO. The computation of NNLO corrections
to such functions is in progress.

\section{Hadroproduction of heavy mesons}\vspace{-0.45cm}
\section{in a massive VFNS}

We had a presentation from I.~Schienbein
on hadroproduction of heavy mesons in a
Massive Variable Flavour Number Scheme (MVFNS). 
Considering, for example, 
$D$-meson production at the Tevatron ($p\bar p\to DX$), 
the MVFNS subtracts and 
resums logarithms of the heavy-quark mass $\ln(\mu_F^2/m^2)$. Unlike
the perturbative fragmentation function approach discussed above, it 
keeps powers of $m/Q$ in the hard-scattering cross section.
This prescription is equivalent to $\overline{\mathrm{MS}}$ mass 
factorization in a scheme where the heavy-quark mass regularizes the
collinear divergences. It was numerically checked that the
short-distance coefficient function corresponds to the 
$\overline{\mathrm{MS}}$ one in the limit where the heavy-quark mass tends to 
zero.
As a result, predictions can be obtained still using
$\overline{\mathrm{MS}}$ parton distributions and fragmentation 
functions, but convoluted with a massive hard-scattering cross section.

The considered partonic subprocesses that were calculated
are $gg\to c\bar c$ and 
$q\bar q\to c\bar c$, at LO; $qq\to c\bar cg$, $q\bar q\to c\bar c g$
and $gq\to c\bar cq$, at NLO.
If the heavy (charm) quark is in the initial state,
it is treated in the massless approximation.

Using the CTEQ6M parton distribution function set and the 
Binnewies--Kniehl--Kr\"amer (BKK) NLO fragmentation functions, fitted to the
OPAL data, predictions were given on the transverse momentum of 
$D^0$, $D^{*+}$, $D^+$ and $D_s^+$ mesons at the Tevatron. 
It was considered just prompt charm production, while $D$ production from
$B$-meson decays was not taken into account.
Within the error range, agreement was found with the CDF data, although
the ratio of the central values of theoretical predictions and data is
about 1.5--1.8, which may warrant further investigation.
A prediction was finally given for the transverse momentum distribution
of $c$-flavoured baryons $\Lambda^+_c$.

This approach will be in the near future extended to $B$-meson production
at the Tevatron, and both $D$ and $B$ mesons in Deep Inelastic Scattering.

\section{Heavy-quarkonium production}

The talk by J.--P. Lansberg discussed the production of 
heavy quarkonium ($Q\bar Q$) in a new model, which consists of an extension
of the Colour Singlet Model (CSM) and the Colour Octet Model (COM).
The naive CSM factorizes heavy-quarkonium production in a hard and a soft part.
In the hard process,  
$Q$ and $\bar Q$ are assumed to be on-shell, in a
colour-singlet state, with zero relative momentum, in a $^3S_1$ 
angular-momentum state (for $J/\psi$, $\psi^\prime$ and $\Upsilon$).
As far as the soft part is concerned, 
the amplitude for the quark-binding probability
is a  wave function, solution of the Schr\"odinger equation.
This model is, however, unable to reproduce the Tevatron Run I data from CDF on
$J/\psi$ and $\psi^\prime$ direct production.

The COM proposes instead 
that quarkonium states are produced by the fragmentation of a
gluon, which is transversely polarized, so that, according to Non-Relativistic
QCD (NRQCD), the quarkonium is to have itself transverse polarization.
Nevertheless, this prediction disagrees with CDF measurements of
unpolarized or slightly longitudinally polarized quarkonium states.

The new model
goes beyond the static and on-shell approximations of the CSM.
$^3S_1$ quarkonium (${\cal Q}=Q\bar Q$) is produced via gluon fusion
$gg\to {\cal Q}g$, but it includes new contributions with respect
to the usual CSM (see J.--P.~Lansberg's presentation in
these proceedings for details).

To describe the soft, non-perturbative part, two  
phenomenological vertex functions are chosen:
$\psi(\vec p_{\mathrm{rel}})\sim \exp[-\vec p_{\mathrm{rel}}^2/\Lambda^2]$
and $\psi(\vec p_{\mathrm{rel}})\sim 
\left(1+\vec p^2_{\mathrm{rel}}/\Lambda^2\right)^{-2}$,
where $\vec p_{\mathrm{rel}}$ is the relative $Q\bar Q$ momentum and
$\Lambda$ is a free size parameter.
The results yielded by this model 
are in good agreement with RICH data on $J/\psi$,
and Tevatron data on $J/\psi$, $\psi^\prime$ and $\Upsilon(1S)$ production.
Fragmentation contributions are then taken from the COM, which gives 
transverse polarization, and included in the model.
The agreement with polarization measurements at the Tevatron is 
shown to be now pretty good.

\section{NLO charmonium production in $\gamma\gamma$ collisions}

The talk given by B.~Kniehl
discussed NLO charmonium production in
photon--photon collisions. The computation of NLO corrections to such processes
is a relevant improvement, since it 
reduces renormalization and factorization scale dependence and allows 
a test of NRQCD.

New results were presented for processes 
$\gamma\gamma\to J/\psi X$, with direct photons and prompt $J/\psi$'s, 
within the framework of NRQCD. In the considered processes,
$X$ can be a purely hadronic state, or a hadronic state with a prompt
photon. Phenomenological results were presented for $e^+e^-$ colliders,
with characteristics similar to the TESLA Linear Collider
project, and a centre-of-mass energy
$\sqrt{s}=500$~GeV. Cuts on transverse momentum and rapidity of
final-state photons were set to $p_T^\gamma>3$~GeV and $|y^\gamma|<2.79$.

It was shown that, unlike the LO, the  NLO prediction yields very little
dependence on the phase-space slicing parameters adopted to cancel soft
and collinear singularities.
Results were presented for the $J/\psi$ rapidity and transverse momentum 
spectrum at LO and NLO, along with the NLO $K$-factor. NLO
corrections exhibit a remarkable impact on both shape and normalization of
the distributions which were shown.

In fact, for $\gamma\gamma\to J/\psi X$, the $K$ factor is large because
of the subprocesses $\gamma\gamma\to c\bar c [^3S_1^{(8)}]g$ 
and $\gamma\gamma\to c\bar c [^3S_1^{(8)}]q\bar q$, which are
mediated by the gluon splitting $g\to c\bar c$, where the $c\bar c$ pair
is in a $^3S_1^{(8)}$ state.

The presented analysis will be extended to electron--proton 
photoproduction and hadroproduction, which will yield
predictions that could be compared with data from HERA II, the
Tevatron Run II and, ultimately, the LHC.

\section{Conclusions}

The heavy-flavour session of DIS 2005 had a number
of interesting theory talks.

The reported progress in heavy-quark parton distribution functions
will be a key ingredient for precision studies of heavy-flavour physics
at present and future high-energy colliders, such as the LHC.

Large-$x$ resummation in the coefficient function for
heavy-quark production in DIS will allow us to extract resummed parton
densities as well.
The NNLO calculation of the initial condition of 
heavy-quark perturbative fragmentation functions could allow the promotion
of the perturbative fragmentation function approach to NNLO/NNLL accuracy if
time-like Altarelli--Parisi splitting functions were to be known 
at NNLO. Given the process independence of the perturbative fragmentation
function, we shall be able to apply 
such results to any process whose coefficient
functions are known to NNLO.

An alternative approach to address heavy-quark (hadron) production is 
the MVFNS, which was used to predict 
charm-flavoured hadron production at the Tevatron. Some discrepancies 
between theory and CDF data in the central values of the $D$ 
transverse-momentum distribution may require further investigation.
It may also be worthwhile
comparing MVFNS and perturbative fragmentation function approaches.

Quarkonium production studies were also discussed.
A new model was proposed, which goes beyond the static approximation
of the CSM, uses the COM for the fragmentation, and is able to
reproduce Tevatron and RICH data on $J/\psi$, $\psi^\prime$ and
$\Upsilon$ production. 

NLO charmonium production in two-photon collisions
was also discussed. The results, shown for $J/\psi$
production at an $e^+e^-$ collider with 
$\sqrt{s}=500$~GeV, exhibit a relevant effect of the inclusion of NLO
corrections.

In summary, all the given talks show active work and progress on heavy-flavour
phenomenology, which will allow the performance of increasingly more accurate 
measurements in Deep Inelastic Scattering experiments, as well as at
any present and future hadron collider facility.










\end{document}